\documentclass[aps,prd,twocolumn,noshowpacs,superscriptaddress,groupedaddress, floatfix,superscriptaddress, preprintnumbers]{revtex4-1}
\usepackage{epsfig}
\usepackage{amsmath}
\usepackage{amsfonts}
\usepackage{amssymb}
\usepackage{graphicx}
\usepackage{color}
\usepackage{slashed}
\usepackage[breaklinks,colorlinks=true]{hyperref}
\usepackage{multirow}
\usepackage{float}
\usepackage[T1]{fontenc}
\usepackage[latin2]{inputenc}%
\begin{document}

\title{Geometric Scaling and the Odderon}

\author{Micha\l~Prasza\l owicz}
\email{michal.praszalowicz@uj.edu.pl \\ \,}
\affiliation{Institute of Theoretical Physics, Jagiellonian University, S. \L ojasiewicza 11, 30-348 Krak\'ow, Poland}

\date{\today}

\begin{abstract}
 Based on geometric scaling of elastic $pp$ and possibly $p\bar{p}$ cross sections, we derive simple formulae for
  the complex crossing-even and crossing-odd scattering amplitudes in terms of two interaction radii: $R(s)$ and $Q(s)$, respectively.
  Employing the COMPETE parametrization of the total cross sections, we reproduce $\rho^{pp}$ and $\rho^{p\bar{p}}$ parameters with high 
  accuracy, however we miss the 13~TeV TOTEM and ATLAS points. We show that a slight modification of the $p\bar{p}$ total cross section
  corresponding to the Odderon, allows to accommodate $\rho^{pp}$ 13~TeV data. 
\end{abstract}

\maketitle

\section{Introduction}

In 2017, the TOTEM Collaboration at CERN released two papers \cite{TOTEM:2017asr,TOTEM:2017sdy} reporting the measurements of the
total $pp$ cross section and the $\rho^{pp}$ parameter (defined as $\rho^{pp}={\rm Re} T_{\rm el}^{pp}(s,t=0)/{\rm Im} T_{\rm el}^{pp}(s,t=0)$) at the highest LHC
energy of 13~TeV. While the total cross section of $110.6 \pm 3.4$~mb was within the expectations based on the COMPETE \cite{Cudell:2001pn} 
parametrization \cite{PDG2010}, the value of the
$\rho^{pp}$ parameter was unexpectedly low: $0.098 \pm 0.01$. In 2023 ATLAS Collaboration \cite{ATLAS:2022mgx} confirmed this result, although the total cross section
turned out to be about $5.8$ mb lower:
$104.7\pm 1.1$~mb. One should remark, however, that various authors have scrutinized TOTEM data analysis, obtaining 
higher values of  $\rho$ \cite{Pancheri:2018yhd,Ezhela:2020hws}.

These results triggered once again speculations that the difference between the COMPETE prediction and the experimental  value of $\rho^{pp}$ is due to the Odderon
exchange \cite{Martynov:2017zjz}. Odderon corresponds to the $C$-odd amplitude that is different for $pp$ and $p\bar{p}$ scattering. 
Introduced in 1973 \cite{Lukaszuk:1973nt}, Odderon is today understood as an exchange of three reggeized gluons coupled symmetrically 
to the SU(3) color singlet \cite{Bartels:1980pe,Kwiecinski:1980wb}.
At lower energies $C$-odd exchanges are due to the $\rho$ and $\omega$ Reggeons, which, however, become negligible at the LHC.

The situation in the literature is confusing. On the one hand, some authors, based on theoretical models of
differential cross sections or on model independent analyses
argue that they are able to describe the data without introducing the Odderon
\cite{Khoze:2017swe,Shabelski:2019alw},
others believe that this is essentially impossible or at least difficult
\cite{Broilo:2018els,Broilo:2018qqs,Broilo:2019xhs},
while there are also reports that the Odderon has undoubtedly been discovered or at least not excluded
\cite{Martynov:2017zjz,Martynov:2018sga,Martynov:2019jwb,TOTEM:2018psk,Csorgo:2018uyp,Csorgo:2019ewn,Csorgo:2020wmw,D0:2020tig,Gotsman:2020mkd,Cui:2022dcm,Csorgo:2024dvr,
Selyugin:2024ccc,velazquezCorral:2026gex}.
A critical review of the existence of the Odderon can be found in \cite{Ryskin:2024qpq,Petrov:2022stg}, see also \cite{Grafstrom:2023hql}.

The approach we present in this article is a modest one. Our goal is to determine whether the forward $pp$ and $p\bar{p}$ 
data allow for, or require, the introduction of the Odderon.
One has to remember that the $\rho^{pp}$ parameter and the total cross section are related, as $\sigma_{\rm tot}^{pp}\sim {\rm Im} T^{pp}_{\rm el}(s,t=0)$,
and ${\rm Im} T^{pp}_{\rm el}(s,t=0)$ in turn  is related through analyticity and dispersion relations to ${\rm Re} T^{pp}_{\rm el}(s,t=0)$. Nevertheless,
implementing these relationships in practice is not straightforward. However, already in the seventies, an effective method  for constructing 
even under crossing, complex amplitudes was proposed in Refs.~\cite{DiasdeDeus:1975ybq,DiasdeDeus:1976em} in the context of geometric
scaling at the ISR \cite{DiasDeDeus:1973lde,Buras:1973km}. We recently used this approach \cite{Praszalowicz:2025twy} 
to study scaling of the elastic cross section
at the LHC \cite{Baldenegro:2022xrj,Baldenegro:2024vgg}. Here we extend this method to amplitudes odd under crossing symmetry and
argue that at the ISR energies they correspond to the Reggeon exchanges, whereas at the LHC a new contribution corresponding to the Odderon is required. 
We show that adding the Odderon to the standard parametrizations of the total cross sections can substantially modify the $\rho$ parameters leaving
the total cross sections almost unchanged.

\section{Geometric scaling} Geometric scaling (GS) follows from two observations, namely that the ratio 
of  bump to  dip {\em positions}, $t_{\rm bump}/t_{\rm dip}$, of the differential elastic $pp$
cross sections is energy independent from the ISR to the LHC \cite{Baldenegro:2024vgg}, and also the ratio  $\sigma_{\rm tot}^{pp}/ \sigma_{\rm el}^{pp}$ 
does not depend on energy \cite{DiasDeDeus:1973lde,Buras:1973km}. While the latter is not true at the LHC, the Ansatz for  the
elastic amplitude satisfying the position scaling takes a very simple form
\begin{equation}
{T}^{pp}_{\text{el}}(s,t) =  i sR^{2}(s)\Phi(\tau)
\label{eq:Tel1}
\end{equation}
in the normalization where ${T}^{pp}_{\text{el}}$ is dimensionless and purely imaginary. Here $R^2$ is the so called interaction radius and
$\Phi$ is a universal function of the scaling variable
$\tau=|t R^2(s)|$, to be determined from the data.  It follows from the optical theorem
\begin{equation}
s\sigma_{\text{tot}}^{pp}(s)=2\operatorname{Im}{T}_{\text{el}}(s,0)
\label{eq:optical}%
\end{equation}
that $R^2(s)=\sigma_{\text{tot}}^{pp}(s)/2$, assuming normalization where $\Phi(0)=1$.

\section{Crossing}
Amplitude (\ref{eq:Tel1}) does not have any well defined behavior under the crossing symmetry.   At high energies in 
the Regge limit (all masses equal zero,
$s\gg-t$) one defines crossing even/odd amplitudes as follows%
\begin{equation}
{T}^{\pm}_{\text{el}}(u,t)\simeq {T}^{\pm}_{\text{el}}(-s,t)= \pm
{T}_{\text{el}}^{\pm\, \ast}(s,t). \label{eq:crossing}%
\end{equation}
Amplitudes for $pp$ and $p\bar{p}$
take then the following form
\begin{equation}
{T}_{\mathrm{el}}^{pp}={T}_{\mathrm{el}}^{+} 
+{T}_{\mathrm{el}}^{-},
~~~
{T}_{\mathrm{el}}^{p{\bar{p}}}={T}_{\mathrm{el}}^{+}-{T}_{\mathrm{el}}^{-}.
\label{eq:Tppandppbar}
\end{equation}

The proper Ansatz for the crossing even amplitude, which satisfies (\ref{eq:crossing}), can be found e.g. in Ref.~\cite{DiasdeDeus:1975ybq}
\begin{equation}
{T}^{+}_{\text{el}}(s,\tau)=isR^{2}(-is)\Phi\Big( \left\vert t  \right\vert R^{2}(-is) \Big) .
\label{eq:Tplussdef}
\end{equation}
We now make an analogous Ansatz for the crossing odd amplitude
\begin{equation}
{T}_{\text{el}}^{-}(s,\tau)=-sQ^{2}(-is)\Psi \Big(  \left\vert t \right\vert Q^{2}(-is)   \Big) , 
\label{eq:Tminusdef}%
\end{equation}
where $Q^2$ is a new interaction radius and $\Psi$  is a new scaling function of a new scaling variable, normalized to 1 at $t=0$.
 However, it is also possible -- as recently suggested in Ref.~\cite{velazquezCorral:2026gex} -- that the scaling variable
 entering $\Psi$ is the same as for $\Phi$.
 Note that the signs of the amplitudes in Eqs.~(\ref{eq:Tplussdef}) and (\ref{eq:Tminusdef}) are arbitrary, so only the absolute values of $R^2$ and $Q^2$ can be interpreted as radii. With our present conventions, $R^2$ is guaranteed to be positive by the optical theorem.
What is following next, should be considered as a test of assumption (\ref{eq:Tminusdef}). 

To identify real and imaginary parts of amplitudes (\ref{eq:Tplussdef}) and   (\ref{eq:Tminusdef}) one defines rapidity $y$
\begin{equation}
-i\left(\frac{s}{1\;{\rm GeV}^2}\right)=e^{y-i\pi/2}
\end{equation}
and follows the trick of Ref.\,\cite{DiasdeDeus:1975ybq}
\begin{equation}
R^{2}(-is) \rightarrow R^{2}\left(  y-i\frac{\pi}{2}\right)  \simeq
R^{2}(y)-i\frac{\pi}{2}\frac{dR^{2}(y)}{dy} \, \label{eq:expansionR}%
\end{equation}
and analogously for $Q^2(-is)$. We obtain \cite{Praszalowicz:2025twy}
\begin{align}
{T}_{\text{el}}^{+}(s,\tau)  &  =s\left[  iR^{2}\Phi+\frac{\pi}{2}%
\frac{dR^{2}}{dy}\frac{d}{d\tau}\left(  \tau\Phi\right)  \right]  ,\nonumber\\
{T}_{\text{el}}^{-}(s,\tau)  &  =s\left[  i\frac{\pi}{2}\frac{dQ^{2}%
}{dy}\frac{d}{d\tau}\left(  \tau\Psi\right)  -Q^{2}\Psi\right]\, ,
\label{eq:Tpmfinal}
\end{align}
and%
\begin{align}
T_{\text{el}}^{pp}(s,\tau)  &  =
s \left[  i\left\{  R^{2}\Phi+\frac{\pi}%
{2}\frac{dQ^{2}}{dy}\frac{d}{d\tau}\left(  \tau\Psi\right)  \right\} \right. \nonumber \\
 & \left. +\left\{  \frac{\pi}{2}\frac{dR^{2}}{dy}\frac{d}{d\tau}\left(  \tau
\Phi\right)  -Q^{2}\Psi\right\}  \right]  ,\nonumber\\
T_{\text{el}}^{p\bar{p}}(s,\tau)  &  =s\left[  i\left\{  R^{2}\Phi-\frac{\pi
}{2}\frac{dQ^{2}}{dy}\frac{d}{d\tau}\left(  \tau\Psi\right)  \right\} \right. \nonumber \\
&\left.  +\left\{  \frac{\pi}{2}\frac{dR^{2}}{dy}\frac{d}{d\tau}\left(  \tau
\Phi\right)  +Q^{2}\Psi\right\}  \right]  .
\label{eq:Tppandppbarfinal}
\end{align}
Note that in \cite{Praszalowicz:2025twy} we neglected the $C$-odd part, which corresponds to $Q^2=0$ in (\ref{eq:Tpmfinal}) and (\ref{eq:Tppandppbarfinal}),
which gives the result of Eq.~(\ref{eq:Tel1}) for the imaginary part of $T_{\text{el}}^{pp}$.
From Eqs.\,(\ref{eq:Tppandppbarfinal}) we can compute  total cross sections
\begin{align}
\sigma_{\mathrm{tot}}^{pp}  &  =2\left\{  R^{2}+\frac{\pi}{2}\frac
{dQ^{2}}{dy}\right\}  ,\nonumber\\
\sigma_{\mathrm{tot}}^{p\bar{p}}  &  =2\left\{  R^{2}-\frac{\pi}{2}%
\frac{dQ^{2}}{dy}\right\}  , \label{eq:sigmatotal}%
\end{align}
and  $\rho$ parameters
\begin{equation}
\rho^{pp}    =\frac{\frac{\pi}{2}\frac{dR^{2}}{dy}-Q^{2}}{R^{2}+\frac{\pi}%
{2}\frac{dQ^{2}}{dy}},~~~
\rho^{p\bar{p}}   =\frac{\frac{\pi}{2}\frac{dR^{2}}{dy}+Q^{2}}{R^{2}%
-\frac{\pi}{2}\frac{dQ^{2}}{dy}},
\label{eq:rhos}
\end{equation}
 which represent the aforementioned relations between $\rho$ and $\sigma_{\rm tot}$.
 As will be explained below, they depend on a single free parameter $q$, which arises when extracting $Q^2$ from the total cross section data.

Again, in \cite{Praszalowicz:2025twy} we used the same formula for
$\rho^{pp}$ but with $Q^2=0$.

\section{Phenomenology}

In order constrain interaction radii $R^{2}$ and $Q^{2}$ we use
 the COMPETE parametrization \cite{Cudell:2001pn}
(quoted in \cite{PDG2010}) for the total cross sections
\begin{equation}
\sigma_{\mathrm{tot}}^{pp/p\bar{p}}(s)   =Z+C\ln^{2}\left(  \frac{s}{s_{0}}\right) +Y_{1}\left(  \frac{s}{s_{1}}\right)  ^{-\eta_{1}} \mp Y_{2}\left(  \frac{s}%
{s_{1}}\right)  ^{-\eta_{2}}\,,
 \label{eq:parPDG}%
\end{equation}
where $Z=35.45$~mb, $C=0.308$~mb, $Y_{1}=42.53$~mb, $Y_{2}=33.34$~mb,
$s_{0}=28.94$~GeV$^{2}$, $s_{1}=1$~GeV$^{2}$ and $\eta_{1}=0.458$, $\eta
_{2}=0.545$. Note that the $\ln^{2}s$ term corresponds to the Froissaron (nonperturbative Pomeron) and two power-like contributions to the
Reggeon exchanges.

Using (\ref{eq:parPDG}) we obtain
\begin{align}
R^2(s) &=\frac{1}{4}\left(  \sigma_{\mathrm{tot}}^{pp}+\sigma_{\mathrm{tot}}^{p\bar{p}}\right) \nonumber \\
 &=\frac{1}{2}\left(  Z+C\ln^{2}\left( \frac{s}{s_{0}}\right)  +Y_{1}\left(  \frac{s}{s_{1}}\right)  ^{-\eta_{1}}\right), \nonumber \\
\frac{dQ^{2}}{dy}(s)  &  =\frac{1}{2\pi}\left(  \sigma_{\mathrm{tot}}^{pp}-\sigma_{\mathrm{tot}}^{p\bar{p}}\right) =-\frac{1}{\pi}Y_{2}\left(  \frac{s}{s_{1}}\right)  ^{-\eta_{2}} .
\label{eq:inputs1}
\end{align}
Hence
\begin{align}
\frac{dR^{2}}{dy}(s)& =\frac{1}{2}\left(  2C\ln\left(  \frac{s}{s_{0}}\right)
-\eta_{1}Y_{1}\left(  \frac{s}{s_{1}}\right)  ^{-\eta_{1}}\right)  , \nonumber \\
Q^{2}(s)  &  = \frac{1}{\pi\eta_{2}}Y_{2}\left(  \frac{s}{s_{1}}\right)  ^{-\eta_{2}%
}\,+q,
\label{eq:inputs2}
\end{align}
where the integration constant $q$  is to be determined from data.

\begin{figure}[h]
\centering
\includegraphics[width=8cm]{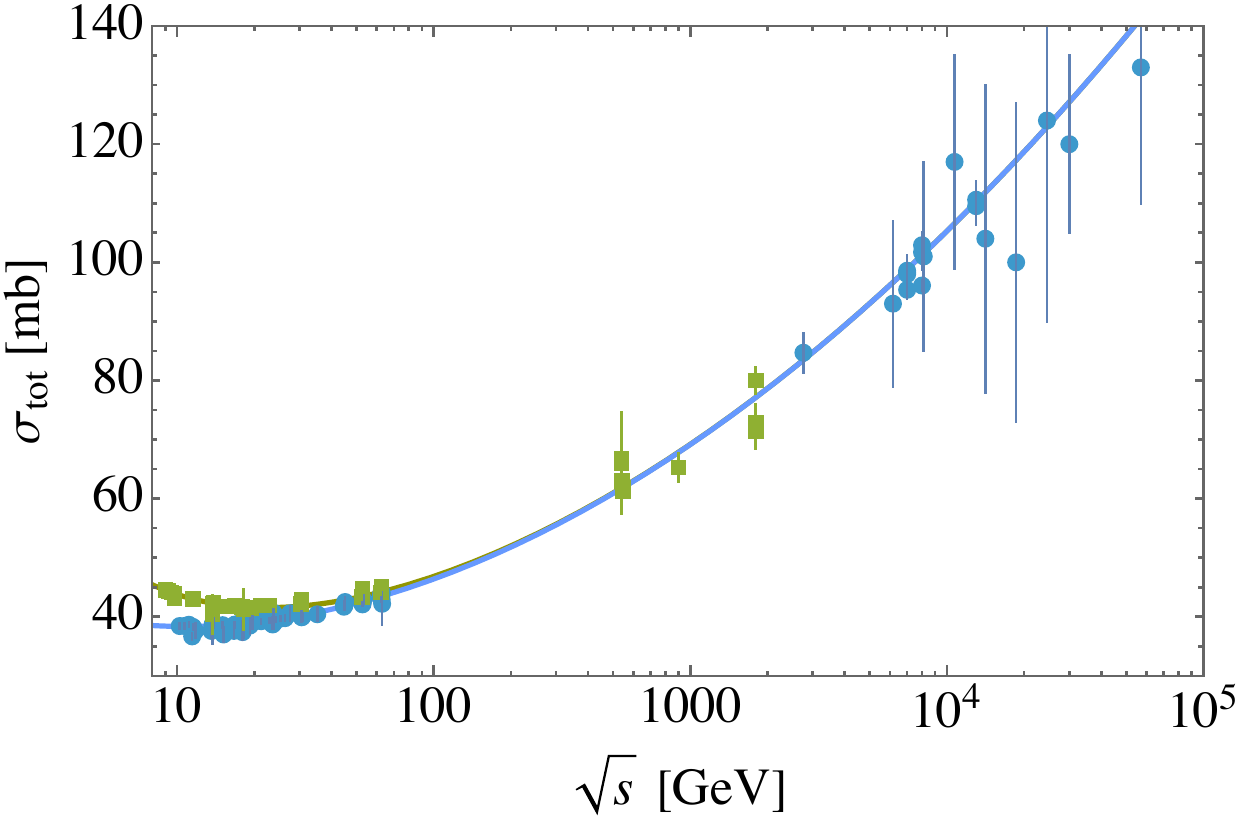} \\
\vspace{0.5cm}
\includegraphics[width=8cm]{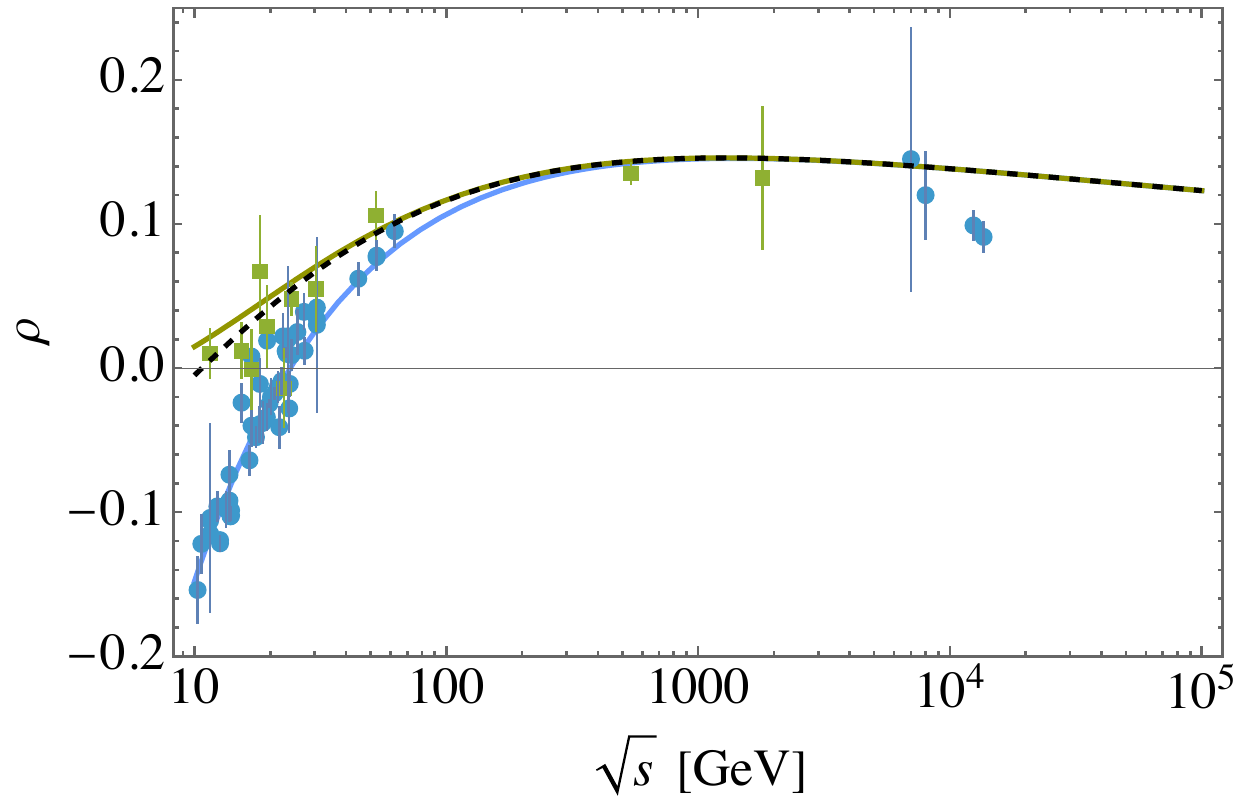}
\vspace{-0.2cm} \caption{\small Upper panel -- total cross sections, lower  panel -- $\rho$ parameter from Eq.~(\ref{eq:rhos})
with $q=0$ (c.f. Eq.~(\ref{eq:inputs2})).
Blue disks $pp$, green squares $p\bar{p}$,
solid lines COMPETE parametrization (\ref{eq:parPDG}).
The black dashed line in the lower panel corresponds to the only  $C$-even parameterization 
from \cite{Praszalowicz:2025twy}. Data from \cite{data}.}%
\label{fig:tor_rho}%
\end{figure}

Now we have all information to compare our  predictions with data. In the upper panel of Fig.~\ref{fig:tor_rho} we plot total $pp$
and $p\bar{p}$ cross sections together with the COMPETE parametrization (\ref{eq:parPDG}). In the lower panel we plot
$\rho^{pp/p\bar{p}}$ given by Eq.~(\ref{eq:rhos}) with $q=0$. We see that  theoretical curves describe data very well. 
On the same plot
we show prediction from Ref.~\cite{Praszalowicz:2025twy}, where $C$-odd amplitude was neglected. We see that inclusion
of a nonzero ${T}_{\text{el}}^{-}$ is absolutely necessary to describe the data with high accuracy. 
This result demonstrates that the parametrization (\ref{eq:Tminusdef}) is a highly plausible possibility. 
However, it tests the crossing-odd contribution and its assumed energy dependence, but not its scaling with momentum transfer.
Obviously, it is related to the $C$-odd Reggeons, and not to the Odderon,
because for parametrization (\ref{eq:parPDG}) the difference
$\sigma_{\mathrm{tot}}^{pp}-\sigma_{\mathrm{tot}}^{p\bar{p}}$ vanishes at high energy (\ref{eq:inputs1}).
Moreover, it still misses the highest energy TOTEM points.

\section{Odderon}

It is clear from the preceding discussion that, in order to reproduce the measured value of  $\rho^{pp}$ at 
13 TeV, a modification of $\sigma^{pp}_{\rm tot}$ or/and $\sigma^{p\bar{p}}_{\rm tot}$
parametrizations 
at high energies is necessary. We decided to modify the $p\bar{p}$ cross section
because it had only been measured up to $\sim 2$~TeV. 
Let us note that there is some freedom in the definition of $R^2$ and $Q^2$ in Eq.~(\ref{eq:sigmatotal}) 
that leaves $\sigma^{pp}_{\rm tot}$ unchanged

\begin{align}
\frac{dQ^{2}(y)}{dy} &  \rightarrow \frac{dQ^{2}(y)}{dy}-\frac{1}{2\pi}\frac{df(y)}{dy}, \nonumber\\
R^{2}(y) &  \rightarrow R^{2}(y)+\frac{1}{4}\frac{df(y)}{dy} ,
\label{eq:gauge}
\end{align}
but modifies $\sigma^{p\bar{p}}_{\rm tot}$
\begin{equation}
\sigma_{\mathrm{tot}}^{p\bar{p}}
=2\left(  R^{2} -\frac{\pi}{2} \frac{dQ^{2}}{dy} \right) +\frac{df}{dy} \, ,
\label{eq:modifiedtotppbar}
\end{equation}
and both $\rho$ parameters
\begin{align}
\rho^{pp}  & =\frac{\frac{\pi}{2}\frac{dR^{2}}{dy} -Q^{2}+
\left(\frac{\pi}{8}\frac{d^{2}f}{dy^{2}}+\frac{1}{2\pi}f \right)}
{R^{2}(y)+\frac{\pi}{2}\frac{dQ^{2}}{dy}}, \nonumber\\
\rho^{p\bar{p}}  & =
\frac{
\frac{\pi}{2}\frac{dR^{2}}{dy}+Q^{2}+ \left( \frac{\pi}{8}\frac{d^{2}f}{dy^{2}}-\frac{1}{2\pi}f \right)}
{\left( R^{2}-\frac{\pi}{2}\frac
{dQ^{2}}{dy}\right)+\frac{1}{2}\frac{df}{dy}}\, ,
\label{eq:modifiedrhos}
\end{align}
where for brevity the $y$ dependence has been suppressed.

The question is whether it is possible to find a function $f(y)$ that only slightly modifies $\sigma^{p\bar{p}}_{\rm tot}$, 
changing at the same time the high energy
part of $\rho^{pp}$. Inspired by the maximal Odderon hypothesis, we propose the following 
Ansatz
\begin{equation}
f(y)=-A \ln^2 (s/s_3)
\label{eq:f}
\end{equation}
with two free parameters $A$ and $s_3$. 

The maximal Odderon hypothesis states that the real part of the $C$-odd amplitude
depends on energy as $\ln^2 s$ (like the imaginary part of the $C$-even amplitude) and was in fact introduced already in 1973
\cite{Lukaszuk:1973nt}. As a consequence the difference between $pp$ and $p\bar{p}$ total cross sections grows with
energy like $\ln s$, see Eq.~(\ref{eq:modifiedtotppbar}). The maximal Odderon hypothesis was invoked to describe the TOTEM $\rho^{pp}$ data
in Refs.~\cite{Martynov:2017zjz,Martynov:2018sga,Martynov:2019jwb}. Even though its theoretical foundations are questioned in the 
literature \cite{Petrov:2022stg,Khoze:2018bus,Petrov:2020mwj}, ,,this does not mean that one cannot use the maximum Odderon parameterization within some limited energy
range'' (quote from \cite{Ryskin:2024qpq}).

The maximal Odderon conjecture has been criticized on the grounds that it may be incompatible with unitarity.  
Although the maximal Odderon does not violate the Froissart bound, since the 
$\log^2 s$ 
contribution to the real part of the scattering amplitude does not contribute directly to the total cross section, it does lead to different coefficients of the  
$\log s$ 
terms in the total cross sections for $pp$ and $p\bar p$ scattering. Consequently, the difference between these cross sections does not vanish asymptotically, 
$\Delta\sigma_{\rm tot} \nrightarrow 0$. Nevertheless, the ratio $\sigma^{pp}_{\rm tot}/\sigma^{p\bar{p}}_{\rm tot} \rightarrow 1$ 
still holds,  representing a generalized form of the Pomeranchuk theorem  \cite{Truong:1972wpr}. The original, stronger statement,  
$\Delta\sigma_{\rm tot} \rightarrow 0$, is not satisfied, but its proof relies on the assumption of non-growing cross sections, 
which is not satisfied at the LHC energies.

Another argument concerning the unitarity of the Odderon amplitude relies on the assumption of black-disk saturation
in the impact parameter space \cite{Khoze:2018oac}, for which the 
$S$-matrix approaches zero asymptotically, $S(s,b) \rightarrow 0$ for each $b$.
This implies that the normalized elastic amplitude approaches $\tilde T_{\rm el}(s,b)\to i$, so that its real part vanishes. 
However, scattering need not become completely absorptive at every impact parameter. It is possible for the local unitarity condition 
to remain satisfied without the $S$-matrix approaching zero, even if the total cross section saturates its asymptotic growth bound \cite{Troshin:2013hia}.
Therefore, the incompatibility of the maximal Odderon with black-disk asymptotics should be distinguished from a general proof 
of its incompatibility with unitarity.

Finally, let us emphasize that the maximal Odderon is a hypothesis of maximal asymptotic growth, corresponding to the saturation 
of the relevant upper bound for the total cross section, rather than a prediction of perturbative QCD. In perturbative QCD, 
the leading-order Odderon intercept 
is equal to unity, whereas the behavior of the nonperturbative Odderon at asymptotically high energies remains uncertain. In particular, 
it is not known whether the nonperturbative Odderon exhibits maximal asymptotic growth or instead behaves like a Regge pole with 
an intercept close to unity. Consequently, the energy range over which the maximal-Odderon hypothesis provides 
a valid phenomenological description cannot presently be determined from first principles.

\section{Numerical results}

At this point one should in principle refit the data including $f$. This, however, is beyond the scope of the present paper
and has been in fact already done in one way or another by various groups. Instead, we keep all fit parameters in (\ref{eq:parPDG})
as they are, 
and search for parameters $q$,  $A$ and $s_3$ that minimize 
mean squared normalized residual
 for the $\rho$ parameters defined as
\begin{equation}
\chi^2=\frac{1}{n_f-n_i+1}\sum_{n=n_i}^{n_f}\left(\frac{\rho_{\rm th}(W_i)-\rho_{\rm exp}(W_i)}{\delta\rho_{\rm exp}(W_i)}\right)^2
\label{eq:chi2}
\end{equation}
to see if adding the Odderon improves the fit to the data. 
In practice
 we chose the following weighting prescription for the $pp$ and $p\bar{p}$ samples
looking for a minimum of $\chi^2=\chi_{pp}^2+\chi_{p\bar{p}}^2$.
We also compute  mean squared normalized residuals of $\sigma_{\rm tot}^{pp}$ and $\sigma_{\rm tot}^{p\bar{p}}$, but in this case separately
for $pp$ and $p\bar{p}$.

We have at our disposal large sets of data on $\rho^{pp}$ and $\rho^{p\bar{p}}$: 118 and 91 points, respectively \cite{data}.
However, most of these data points are at low energies and have large uncertainties. Therefore in computing $\chi^2$
we have used only high energy data starting from $\sqrt{s}=24.3$~{GeV} both for $pp$ and $p\bar{p}$ \cite{UA6:1988mkw}
where the $\rho$ parameters are positive.
This leaves 13 (15) $pp$ data points without (with) TOTEM 13 TeV points and, unfortunately, only 5 $p{\bar{p}}$ points.

In principle, we should have computed $\chi^2$ by varying three parameters: $q$, which is the integration constant for the 
odd $C$ interaction radius in Eq.~(\ref{eq:inputs2}), and the two Odderon parameters, $A$ and $s_3$ (\ref{eq:f}). 
However, since -- as we have checked explicitly -- the dependence on $s_3$ is very weak, we keep it fixed at $s_3=1000$~GeV$^2$. The results are displayed in Table~\ref{tab:fits}.

\begin{table}[h!]
\begin{center}
\begin{tabular}{|c|c|c|c|c|c|c|}
\hline
\# & 13~TeV & $q$~[mb]& $~A$~[mb]~ & ~$\chi_{\rm min}^2$~ & $\chi_{pp}^2$ & $\chi_{p\bar{p}}^2$ \\
\hline
1 & no          &     $0$         &      $0$  &      ${\bf 1.498}$       & ~$0.814 $~    &   ~$ 0.683 $~         \\
2 & no           & $-0.16$      &       $0$ &   ${\bf 0.781} $         &     $ 0.423 $             &      $0.359 $    \\
3 &  no          & $-0.14$   & $-0.015$&  ${\bf 0.758}$   &   $ 0.445$              &      $0.312 $     \\
\hline
4 & yes         & $0$            &   $0$       &  ${\bf 3.729}$        &      $3.046$              &  $0.683$      \\
5 & yes         & $-0.12$      &   $0$       &  ${\bf 3.355}$        &      $2.264$              &  $0.391$      \\
6 & yes         & $-0.23$      &   $0.068$ &  ${\bf 1.845}$       &      $0.720$              &  $1.117$     \\
\hline
\end{tabular}
\end{center}
\caption{Values of   mean squared normalized residuals $\chi^2$ defined in Eq.~(\ref{eq:chi2}) 
for different modifications of the $\rho$ parameters
with and without 13~TeV TOTEM $\rho^{pp}$ points. COMPETE parametrization (\ref{eq:parPDG}) remains unchanged.}
\label{tab:fits}
\end{table}

\begin{figure}[h]
\centering
\includegraphics[width=8cm]{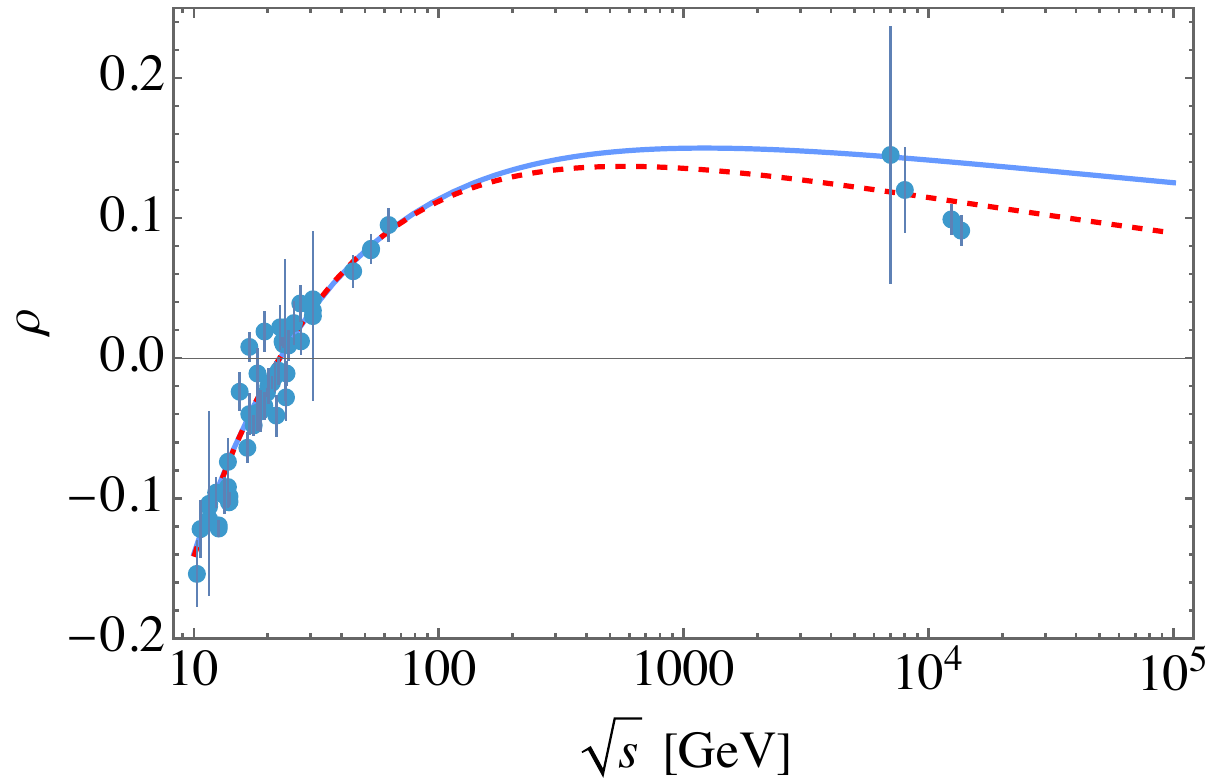}\\
\vspace{0.5cm}
\hspace{0.1cm}
\includegraphics[width=8cm]{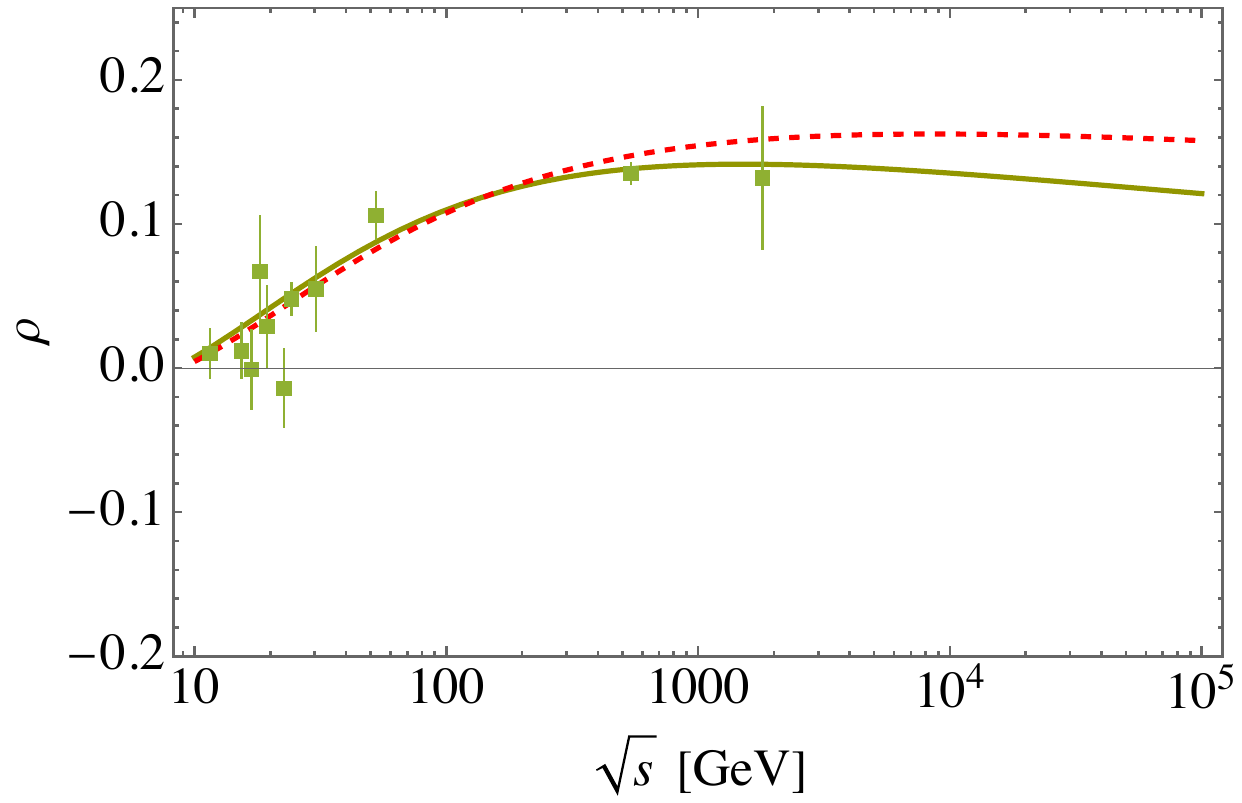}
\vspace{-0.2cm} \caption{\small  $\rho$ parameters for the COMPETE parametrizations
corresponding to fit \# 2 from Tab.~\ref{tab:fits} -- solid line. Modified parametrization
(\ref{eq:modifiedrhos}) with $f$ given by (\ref{eq:f}) corresponding to fit \# 6 from Tab.~\ref{tab:fits}
 -- dashed red lines. Upper panel $pp$, lower panel $p\bar{p}$. Data from \cite{data}.}%
\label{fig:modified_rho}%
\end{figure}

Let us first discuss the case without the 13~TeV TOTEM points. As seen from Tab.~\ref{tab:fits}, the data favor a small integration 
constant, $q=-0.16$~mb, leading to a very good $\chi^2$. Adding the Odderon contribution in Eq.~(\ref{eq:f}) only slightly improves
the $\chi^2$ value. We therefore conclude that the Odderon is not necessary to describe the data below 13~TeV. 
In Figs.~\ref{fig:modified_rho}, fit \#2 is represented by solid lines.

Adding the 13~TeV TOTEM points drastically changes the situation. The value of $\chi^2$ is of the order of 3, 
regardless of the value of the parameter $q$. This change is driven by $\chi^2_{pp}$. Adding the Odderon contribution 
substantially improves $\chi^2_{pp}$ at the expense of $\chi^2_{p\bar{p}}$, which increases 
but remains at an acceptable level. In Figs.~\ref{fig:modified_rho}, fit \#6 is represented by dashed red lines.

Including the ATLAS measurement \cite{ATLAS:2022mgx} of $\rho^{pp}$ at 13 TeV slightly shifts fit parameters:
$A: 0.068 \rightarrow 0.073$ and $q:-0.23\rightarrow -0.24$ improving $\chi^2_{pp}:0.720 \rightarrow 0.661$
at the expense of $\chi^2_{p\bar{p}}:1.117 \rightarrow 1.222$.

\begin{figure}[h]
\centering
\includegraphics[width=8cm]{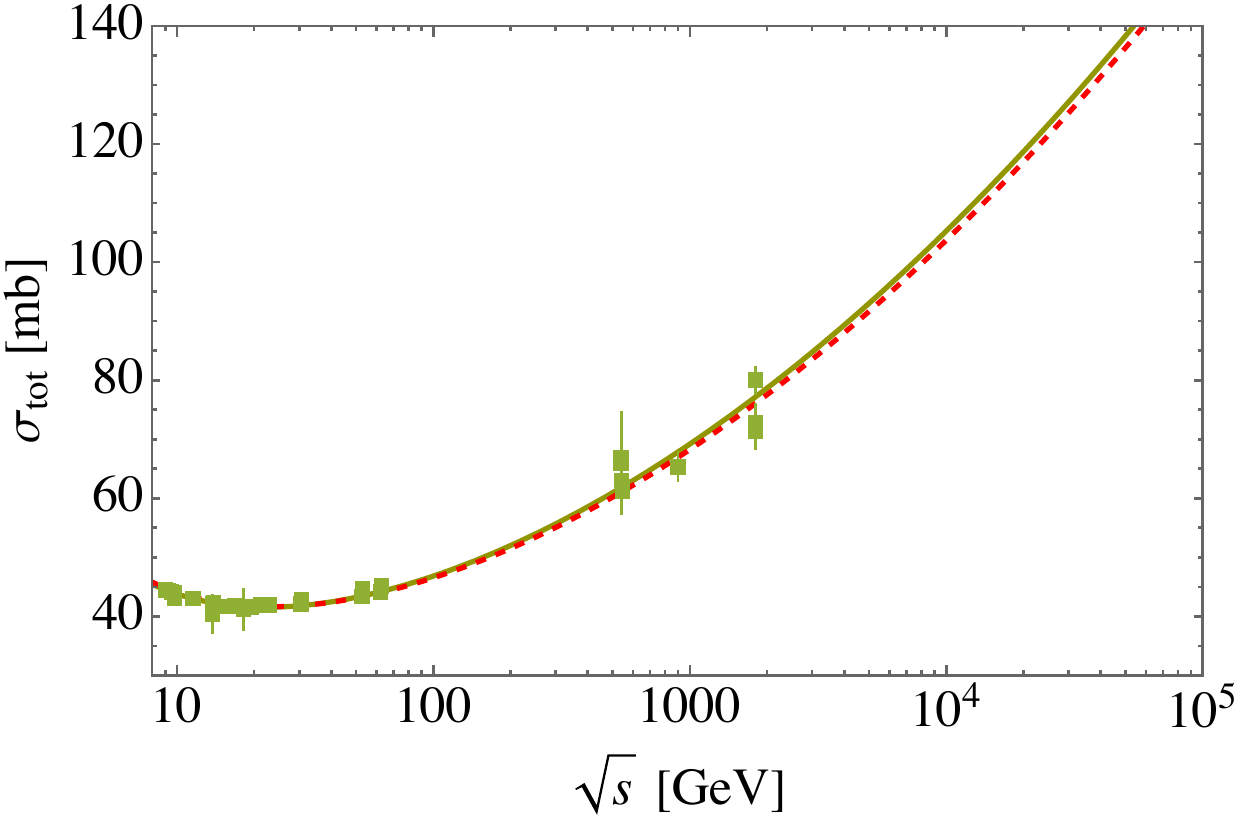} 
\vspace{-0.2cm} \caption{\small  
Total $p\bar{p}$ cross section for the COMPETE parametrization -- solid line, for the
modified case (\ref{eq:modifiedtotppbar}) with $f$ given by  (\ref{eq:f}) 
corresponding to fit \# 6 from Tab.~\ref{tab:fits}
-- dashed red line.
Data from \cite{data}.}%
\label{fig:modified_sigppbar}%
\end{figure}

Finally we have to check how much the Odderon contribution (\ref{eq:f}) modifies numerically the total $p\bar{p}$ cross section.
This is shown in Fig.~\ref{fig:modified_sigppbar}. Let us remark that the COMPETE parametrization alone gives
very good  mean squared normalized residual of $\sigma_{\rm tot}^{p\bar{p}}$, namely $0.736$. Including the Odderon
shifts this value to $0.975$. 

An immediate conclusion from this exercise is that the COMPETE parametrization correctly reproduces $C$-odd amplitudes
at low energies up to the LHC energies below 13~TeV.  
 Keeping the COMPETE parametrization fixed, a modification of its high-energy behavior is required to accommodate the TOTEM 13~TeV data for the $\rho^{pp}$ parameter.
We have shown that the Odderon modification introduced through Eqs.~(\ref{eq:gauge}) and (\ref{eq:f}) provides a very good description of the existing data on both total cross sections and $\rho$ parameters.

\bigskip

\section{Discussion and summary}

In this paper we have concentrated on the forward Odderon signatures. To this end we have used the GS motivated forms of the 
$C$-odd/even scattering amplitudes. Of course to fully appreciate GS and/or the Odderon, one has to investigate the differential 
cross sections, which allows to study
the dip and bump structures in $pp$ and in $p\bar{p}$, along the lines of Ref.~\cite{Praszalowicz:2025twy}. Work on these issues
in the framework of GS
 is currently underway.
Let us also remark that it is not clear whether the Odderon
amplitude $T^{-}_{\rm el}$ does really exhibit GS, although the results of a recent paper \cite{velazquezCorral:2026gex} suggest
that this is most likely the case. However, in the forward kinematics the
impact of the GS factorization hypothesis is minimal.

The simplicity of the relationship between the real and imaginary parts of the scattering amplitude
follows from the expansion (\ref{eq:expansionR}). In principle one could derive such relations
using dispersion relations, which is in many cases impossible to perform analytically
(see however \cite{Avila:2003cu}). We have checked that the higher
order correction to  (\ref{eq:expansionR}) can be safely neglected. This can be seen a'posteriori
from the quality of predictions for $\rho$ parameters at the ISR energies.

While the geometric scaling parametrization of the $C$-even amplitude appears to be a well-established assumption, 
supported by studies such as \cite{DiasdeDeus:1975ybq,DiasdeDeus:1976em,DiasDeDeus:1973lde,Buras:1973km,Praszalowicz:2025twy,Baldenegro:2022xrj},
to the best of our knowledge, geometric scaling of $C$-odd amplitudes is employed for the first time in the present work.
In Ref.~\cite{velazquezCorral:2026gex}, a similar scaling approach was used to study the $C$-odd amplitude, based on fits to the differential cross sections for $pp$ and $p\bar{p}$ scattering.
Similar studies of differential cross sections within the present framework are needed to assess the validity 
of the geometric scaling hypothesis for $C$-odd amplitudes.

With these rather minimalistic assumptions we have been able to reproduce the $\rho$ parameters up to 1~TeV with 
high accuracy. In order to reproduce the TOTEM points we modified $\sigma^{p\bar{p}}_{\rm tot}$ adding the Odderon
contribution (\ref{eq:f}). This was motivated by the fact that $\sigma^{p\bar{p}}_{\rm tot}$  is not constrained above 2~TeV.
Since the correction turned out to be very small, one might just as well modify $\sigma^{pp}$ or both. 
Our study provides further hint of the discovery of the Odderon at the LHC.


\bigskip

\begin{thebibliography}{99}  
\bibitem{TOTEM:2017asr}
G.~Antchev \textit{et al.} [TOTEM],
``First measurement of elastic, inelastic and total cross section at $\sqrt{s}=13$ TeV by TOTEM and overview of cross section data at LHC energies,''
Eur. Phys. J. C \textbf{79} (2019) no.2, 103
doi:10.1140/epjc/s10052-019-6567-0
[arXiv:1712.06153 [hep-ex]].

\bibitem{TOTEM:2017sdy}
G.~Antchev \textit{et al.} [TOTEM],
``First determination of the ${\rho }$ parameter at ${\sqrt{s} = 13}$ TeV: probing the existence of a colourless C-odd three-gluon compound state,''
Eur. Phys. J. C \textbf{79} (2019) no.9, 785
doi:10.1140/epjc/s10052-019-7223-4
[arXiv:1812.04732 [hep-ex]].

\bibitem {Cudell:2001pn}J.~R.~Cudell, V.~Ezhela, P.~Gauron, K.~Kang,
Y.~V.~Kuyanov, S.~Lugovsky, B.~Nicolescu and N.~Tkachenko, ``Hadronic
scattering amplitudes: Medium-energy constraints on asymptotic behavior,''
Phys. Rev. D \textbf{65} (2002), 074024. 


\bibitem {PDG2010}K. Nakamura et al. (Particle Data Group), J. Phys. G 37 (2010),
075021, sec.~41, p.~10.

\bibitem{ATLAS:2022mgx}
G.~Aad \textit{et al.} [ATLAS],
``Measurement of the total cross section and $\rho $-parameter from elastic scattering in pp collisions at $\sqrt{s}=13$~TeV with the ATLAS detector,''
Eur. Phys. J. C \textbf{83} (2023) no.5, 441
doi:10.1140/epjc/s10052-023-11436-8
[arXiv:2207.12246 [hep-ex]].

\bibitem{Pancheri:2018yhd}
G.~Pancheri, S.~Pacetti and Y.~Srivastava,
``Analysis and Implications of precision near-forward TOTEM data,''
Phys. Rev. D \textbf{99} (2019) no.3, 034014
doi:10.1103/PhysRevD.99.034014
[arXiv:1811.00499 [hep-ph]].
\bibitem{Ezhela:2020hws}
V.~V.~Ezhela, V.~A.~Petrov, N.~P.~Tkachenko and A.~A.~Logunov,
``On the $ \rho $ and $ \sigma_{tot} $ measurement by the TOTEM Collaboration: in the wake of recent discoveries,''
[arXiv:2003.03817 [hep-ph]].

\bibitem{Martynov:2017zjz}
E.~Martynov and B.~Nicolescu,
``Did TOTEM experiment discover the Odderon?,''
Phys. Lett. B \textbf{778} (2018), 414-418
doi:10.1016/j.physletb.2018.01.054
[arXiv:1711.03288 [hep-ph]].

\bibitem{Lukaszuk:1973nt}
L.~Lukaszuk and B.~Nicolescu,
``A Possible interpretation of p p rising total cross sections,''
Lett. Nuovo Cim. \textbf{8} (1973), 405-413
doi:10.1007/BF02824484

\bibitem{Bartels:1980pe}
J.~Bartels,
Nucl. Phys. B \textbf{175} (1980), 365-401
doi:10.1016/0550-3213(80)90019-X

\bibitem{Kwiecinski:1980wb}
J.~Kwiecinski and M.~Praszalowicz,
``Three Gluon Integral Equation and Odd c Singlet Regge Singularities in QCD,''
Phys. Lett. B \textbf{94} (1980), 413-416
doi:10.1016/0370-2693(80)90909-0

\bibitem{Khoze:2017swe}
V.~A.~Khoze, A.~D.~Martin and M.~G.~Ryskin,
``Elastic proton-proton scattering at 13 TeV,''
Phys. Rev. D \textbf{97} (2018) no.3, 034019
doi:10.1103/PhysRevD.97.034019
[arXiv:1712.00325 [hep-ph]].
\bibitem{Shabelski:2019alw}
Y.~M.~Shabelski and A.~G.~Shuvaev,
``Unified description of LHC data on elastic $pp$ scattering,''
Mod. Phys. Lett. A \textbf{34} (2019) no.37, 1950305
doi:10.1142/S021773231950305X
[arXiv:1904.07607 [hep-ph]].

\bibitem{Broilo:2018els}
M.~Broilo, E.~G.~S.~Luna and M.~J.~Menon,
``Soft Pomerons and the Forward LHC Data,''
Phys. Lett. B \textbf{781} (2018), 616-620
doi:10.1016/j.physletb.2018.04.045
[arXiv:1803.07167 [hep-ph]].
\bibitem{Broilo:2018qqs}
M.~Broilo, E.~G.~S.~Luna and M.~J.~Menon,
 ``Elastic Scattering and Pomeron Models,''
Phys. Rev. D \textbf{98} (2018) no.7, 074006
doi:10.1103/PhysRevD.98.074006
[arXiv:1807.10337 [hep-ph]].
\bibitem{Broilo:2019xhs}
M.~Broilo, D.~A.~Fagundes, E.~G.~S.~Luna and M.~J.~Menon,
``Forward elastic scattering: Dynamical gluon mass and semihard interactions,''
Eur. Phys. J. C \textbf{79} (2019) no.12, 1033
doi:10.1140/epjc/s10052-019-7545-2
[arXiv:1906.05932 [hep-ph]].


\bibitem{Martynov:2018sga}
E.~Martynov and B.~Nicolescu,
``Odderon effects in the differential cross sections at Tevatron and LHC energies,''
Eur. Phys. J. C \textbf{79} (2019) no.6, 461
doi:10.1140/epjc/s10052-019-6954-6
[arXiv:1808.08580 [hep-ph]].
\bibitem{Martynov:2019jwb}
E.~Martynov and G.~Tersimonov,
%
``Ratio $\rho^{pp}_{\bar pp}(s)$ in Froissaron and maximal odderon approach,''
Phys. Rev. D \textbf{100} (2019) no.11, 114039
doi:10.1103/PhysRevD.100.114039
[arXiv:1911.06873 [hep-ph]].
\bibitem{TOTEM:2018psk}
G.~Antchev \textit{et al.} [TOTEM],
``Elastic differential cross section ${\mathrm{d}}\sigma /{\mathrm{d}}t$ at $\sqrt{s}=2.76\hbox { TeV}$ and implications on the existence of a colourless C-odd three-gluon compound state,''
Eur. Phys. J. C \textbf{80} (2020) no.2, 91
doi:10.1140/epjc/s10052-020-7654-y
[arXiv:1812.08610 [hep-ex]].
\bibitem{Csorgo:2018uyp}
T.~Cs{\"o}rg{\H{o}}, R.~Pasechnik and A.~Ster,
``Odderon and proton substructure from a model-independent L{\'e}vy imaging of elastic $pp$ and $p\bar{p}$ collisions,''
Eur. Phys. J. C \textbf{79} (2019) no.1, 62
doi:10.1140/epjc/s10052-019-6588-8
[arXiv:1807.02897 [hep-ph]].
\bibitem{Csorgo:2019ewn}
T.~Cs{\"o}rg{\H{o}}, T.~Novak, R.~Pasechnik, A.~Ster and I.~Szanyi,
``Evidence of Odderon-exchange from scaling properties of elastic scattering at TeV energies,''
Eur. Phys. J. C \textbf{81} (2021) no.2, 180
doi:10.1140/epjc/s10052-021-08867-6
[arXiv:1912.11968 [hep-ph]].
\bibitem{Csorgo:2020wmw}
T.~Csorgo and I.~Szanyi,
``Observation of Odderon effects at LHC energies: a real extended Bialas{\textendash}Bzdak model study,''
Eur. Phys. J. C \textbf{81} (2021) no.7, 611
doi:10.1140/epjc/s10052-021-09381-5
[arXiv:2005.14319 [hep-ph]].
\bibitem{D0:2020tig}
V.~M.~Abazov \textit{et al.} [D0 and TOTEM],
``Odderon Exchange from Elastic Scattering Differences between $pp$ and $p \bar{p}$ Data at 1.96~TeV and from pp Forward Scattering Measurements,''
Phys. Rev. Lett. \textbf{127} (2021) no.6, 062003
doi:10.1103/PhysRevLett.127.062003
[arXiv:2012.03981 [hep-ex]].
\bibitem{Gotsman:2020mkd}
E.~Gotsman, E.~Levin and I.~Potashnikova,
``New parton model for the soft interactions at high energies: The odderon,''
Phys. Rev. D \textbf{101} (2020) no.9, 094021
doi:10.1103/PhysRevD.101.094021
[arXiv:2003.09155 [hep-ph]].
\bibitem{Cui:2022dcm}
Z.~F.~Cui, D.~Binosi, C.~D.~Roberts, S.~M.~Schmidt and D.~N.~Triantafyllopoulos,
``Fresh look at experimental evidence for odderon exchange,''
Phys. Lett. B \textbf{839} (2023), 137826
doi:10.1016/j.physletb.2023.137826
[arXiv:2205.15438 [hep-ph]].
\bibitem{Csorgo:2024dvr}
T.~Cs{\"o}rg{\H{o}}, T.~Nov{\'a}k, R.~Pasechnik, A.~Ster and I.~Szanyi,
``Model-Independent Odderon Results Based on New TOTEM Data on Elastic Proton{\textendash}Proton Collisions at 8 TeV,''
Universe \textbf{10} (2024) no.6, 264
doi:10.3390/universe10060264
[arXiv:2405.06733 [hep-ph]].
\bibitem{Selyugin:2024ccc}
O.~V.~Selyugin,
``Elastic scattering at s=6{\,}{\,}GeV up to s=13{\,}{\,}TeV: Proton-proton; proton-antiproton, and proton-neutron,''
Phys. Rev. D \textbf{110} (2024) no.11, 114028
doi:10.1103/PhysRevD.110.114028
[arXiv:2407.01311 [hep-ph]].
\bibitem{velazquezCorral:2026gex}
J.~A.~Velazquez Corral, B.~G.~Giraud, R.~Peschanski and C.~Royon,
``Determination of the Odderon amplitude in elastic cross sections at high energies from scaling and analyticity,''
[arXiv:2607.11286 [hep-ph]].


\bibitem{Ryskin:2024qpq}
M.~G.~Ryskin,
``Current Status of the Odderon,''
Phys. Part. Nucl. Lett. \textbf{22} (2025) no.1, 187-190
doi:10.1134/S1547477124702078
[arXiv:2408.01990 [hep-ph]].
\bibitem{Petrov:2022stg}
V.~Petrov and N.~Tkachenko,
``Odderon: Lost or/and Found?,''
[arXiv:2201.06948 [hep-ph]].

\bibitem{Grafstrom:2023hql}
P.~Grafstrom,
``The total cross section for proton-proton interactions at the FCC,''
[arXiv:2306.15449 [hep-ph]].

\bibitem{DiasdeDeus:1975ybq}
J.~Dias de Deus,
``On the Real Part of a Geometrical Pomeron,''
Nuovo Cim. A \textbf{28} (1975), 114
doi:10.1007/BF02730400

\bibitem{DiasdeDeus:1976em}
J.~Dias de Deus and P.~Kroll,
``On the Systematics of Elastic Scattering at High and Intermediate-Energy,''
Nuovo Cim. A \textbf{37} (1977), 67
doi:10.1007/BF02790583

\bibitem{DiasDeDeus:1973lde}
J.~Dias De Deus,
``Geometric Scaling, Multiplicity Distributions and cross sections,''
Nucl. Phys. B \textbf{59} (1973), 231-236
doi:10.1016/0550-3213(73)90485-9

\bibitem{Buras:1973km}
A.~J.~Buras and J.~Dias de Deus,
``Scaling law for the elastic differential cross section in p p scattering from geometric scaling,''
Nucl. Phys. B \textbf{71} (1974), 481-492
doi:10.1016/0550-3213(74)90197-7


\bibitem{Praszalowicz:2025twy}
M.~Prasza{\l}owicz,
``Geometric scaling of elastic pp cross section at the LHC,''
Phys. Lett. B \textbf{869} (2025), 139848
doi:10.1016/j.physletb.2025.139848
[arXiv:2504.18841 [hep-ph]].

\bibitem{Baldenegro:2022xrj}
C.~Baldenegro, C.~Royon and A.~M.~Stasto,
``Scaling properties of elastic proton-proton scattering at LHC energies,''
Phys. Lett. B \textbf{830} (2022), 137141
doi:10.1016/j.physletb.2022.137141
[arXiv:2204.08328 [hep-ph]].

\bibitem{Baldenegro:2024vgg}
C.~Baldenegro, M.~Praszalowicz, C.~Royon and A.~M.~Stasto,
``Scaling laws of elastic proton-proton scattering differential cross sections,''
Phys. Lett. B \textbf{856} (2024), 138960
doi:10.1016/j.physletb.2024.138960
[arXiv:2406.01737 [hep-ph]].

\bibitem{data}
R.~L.~Workman \textit{et al.} [Particle Data Group],
``Review of Particle Physics,''
PTEP \textbf{2022} (2022), 083C01
doi:10.1093/ptep/ptac097,
https://pdg.lbl.gov/2022/hadronic-xsections/hadron.html \, .

\bibitem{Khoze:2018bus}
V.~A.~Khoze, A.~D.~Martin and M.~G.~Ryskin,
``Black disk, maximal Odderon and unitarity,''
Phys. Lett. B \textbf{780} (2018), 352-356
doi:10.1016/j.physletb.2018.03.025
[arXiv:1801.07065 [hep-ph]].

\bibitem{Petrov:2020mwj}
V.~A.~Petrov,
``On the {\textquotedblleft}Froissaron-maximal Odderon{\textquotedblright} model,''
Eur. Phys. J. C \textbf{81} (2021) no.7, 670
doi:10.1140/epjc/s10052-021-09465-2
[arXiv:2008.00990 [hep-ph]].

\bibitem{Truong:1972wpr}
T.~N.~Truong and W.~S.~Lam,
``Proof of the pomeranchuk theorem and related theorems deduced from unitarity,''
Phys. Rev. D \textbf{6} (1972), 2875-2883
doi:10.1103/PhysRevD.6.2875

\bibitem{Khoze:2018oac}
V.~A.~Khoze, A.~D.~Martin and M.~G.~Ryskin,
``Black disk radius constrained by unitarity,''
Phys. Lett. B \textbf{787} (2018), 167-170
doi:10.1016/j.physletb.2018.10.054
[arXiv:1809.10406 [hep-ph]].

\bibitem{Troshin:2013hia}
S.~M.~Troshin and N.~E.~Tyurin,
``Is elastic scattering at the LHC absorptive or geometric?,''
Phys. Rev. D \textbf{88} (2013), 077502
doi:10.1103/PhysRevD.88.077502
[arXiv:1307.0283 [hep-ph]].

\bibitem{UA6:1988mkw}
R.~E.~Breedon \textit{et al.} [UA6],
``Precise Comparison of Anti-proton - Proton and Proton Proton Forward Elastic Scattering at $\sqrt{s}=24.3$-{GeV},''
Phys. Lett. B \textbf{216} (1989), 459-465
doi:10.1016/0370-2693(89)91150-7

\bibitem{Avila:2003cu}
R.~F.~Avila and M.~J.~Menon,
``Critical analysis of derivative dispersion relations at high-energies,''
Nucl. Phys. A \textbf{744} (2004), 249-272
doi:10.1016/j.nuclphysa.2004.08.014
[arXiv:hep-ph/0309028 [hep-ph]].



\end{thebibliography}
\end{document}